\documentclass[prd,
preprint,
superscriptaddress,
preprintnumbers,
eqsecnum,
showpacs,
nofootinbib,
nobibnotes]{revtex4}
\usepackage{amsfonts,bm,amsmath}
\usepackage{graphicx}
\usepackage{color}

\newcommand{\be}{\begin{equation}}
\newcommand{\bea}{\begin{eqnarray}}
\newcommand{\beq}{\begin{eqnarray}}
\newcommand{\ee}{\end{equation}}
\newcommand{\eea}{\end{eqnarray}}
\newcommand{\eeq}{\end{eqnarray}}

\def\s#1{{\scriptscriptstyle #1}}
\def\eq#1{Eq.~(\ref{#1})}

\def\1eq#1{Eq.~(\ref{#1})}
\def\2eqs#1#2{Eqs.~(\ref{#1}) and~(\ref{#2})}
\def\3eqs#1#2#3{Eqs.~(\ref{#1}), (\ref{#2}) and~(\ref{#3})}
\def\4eqs#1#2#3#4{Eqs.~(\ref{#1}), (\ref{#2}), (\ref{#3}) and~(\ref{#4})}
\def\noeq#1{(\ref{#1})}

\def\fig#1{Fig.~\ref{#1}}

\def\ie{{\it i.e.}, }
\def\eg{{\it e.g.}, }

\def\nf{N_f}
\def\nc{N_\s{C}}

\newcommand{\re}{\text{Re}}
\newcommand{\tr}{\text{Tr}}

\begin{document}

\title{Quark flavour effects on gluon and ghost propagators}

\author{A.~Ayala}
\affiliation{Dpto. F\'isica Aplicada, Fac. Ciencias Experimentales;
Universidad de Huelva, 21071 Huelva; Spain.}
\author{A.~Bashir}
\affiliation{Instituto de F\'{\i}sica y Matem\'atica, Universidad Michoacana de San Nicol\'as
de Hidalgo; Edificio C-3, Ciudad Universitaria, Morelia,
Michoac\'an 58040, Mexico.}  \affiliation{ Physics
Division, Argonne National Laboratory, Argonne, Illinois 60439,
USA.} \affiliation{Center for Nuclear
Research, Department of Physics, Kent State University, Kent, Ohio
44242, USA.}
\author{D.~Binosi}
\affiliation{European Center for Theoretical Studies in Nuclear Physics and Related Areas (ECT*) and
Fondazione Bruno Kessler; Villa Tambosi, Strada delle Tabarelle 286, I-32123 Villazzano (TN) Italy}
\author{M.~Cristoforetti}
\affiliation{European Center for Theoretical Studies in Nuclear Physics and Related Areas (ECT*) and
Fondazione Bruno Kessler; Villa Tambosi, Strada delle Tabarelle 286, I-32123 Villazzano (TN) Italy}\affiliation{LISC, Via Sommarive 18, Povo (Trento), I-38123  Italy}
\author{J.~Rodr\'{\i}guez-Quintero}
\affiliation{Dpto. F\'isica Aplicada, Fac. Ciencias Experimentales;
Universidad de Huelva, 21071 Huelva; Spain.}

\begin{abstract}

We compute the full non-perturbative ghost and gluon two-point
Green functions by using gauge field configurations with $\nf=2$
and $\nf=2+1+1$ twisted-mass quark flavours. We use simulations
with several different light quark masses,  heavy quark
masses close to that of the strange and charm quarks, and lightest
pseudoscalar masses ranging from $270$~to $510$~[MeV].  Quark
 flavour effects on both the gluon and the ghost
propagators  are then investigated  in a
wide range of momenta,  bridging the  deep
infrared and intermediate momenta domain of QCD interactions in
the presence of dynamical quarks.  The ghost-gluon
vertex is also indirectly probed through a consistency requirement
among the lattice data for the gluon and ghost propagators and the
ghost propagator Schwinger-Dyson equation. The effective full QCD
coupling is finally constructed, and its dependence on the
presence of dynamical fermions scrutinized. 
\end{abstract}

%\begin{flushleft}
%\begin{figure}[!h]
%  \begin{center}
%    \includegraphics[scale=1]{ETMC_rund.pdf}
%  \end{center}
%\end{figure}
%\end{flushleft}


\maketitle

\section{Introduction}

During the past few years, lattice simulations have considerably
improved our understanding of the infrared (IR) sector of
non-abelian Yang Mills theories. In particular, {\it quenched}
Landau gauge simulations~\cite{Cucchieri:2007md,Cucchieri:2010xr,
Bogolubsky:2007ud,Bogolubsky:2009dc,Oliveira:2009eh}, performed on
lattices with large volumes, have unequivocally demonstrated that
the gluon propagator saturates in the (deep) IR region. This is
true for the space-time dimensions $d=3$ as well as 4,
irrespectively of the number of colors $\nc$ of the gauge group
$SU(\nc)$ under consideration. At the same time, the ghost
dressing function effectively acquires its tree-level behaviour,
with the functional form of the propagator being $\sim 1/q^2$.

Within the continuum formulation of the theory, these
lattice results are in agreement with the solutions of the
corresponding all-order Schwinger-Dyson equations
(SDEs)~\cite{Aguilar:2008xm,Boucaud:2008ky}  and exact
renormalization group (RG) equations~\cite{Fischer:2008uz}. Other
approaches such as the so-called refined Gribov-Zwanziger
formalism~\cite{Dudal:2008sp} also converge to the same
conclusions.  This has caused a paradigmatic shift among
practitioners: the gluon is now thought to acquire a
momentum-dependent mass $m(q^2)$ whose magnitude can be large at
IR momenta,
%$m^2_g(k^2 \sim 0) \simeq (2-4 \Lambda_{\rm QCD})^2$. However, it
but vanishes with increasing spacelike momenta (\ie $q^2\gg
\Lambda_\s{\rm QCD}^2$), thereby maintaining full accord with
perturbative QCD. 
Gluon confinement is then realized, as it is customarily done in the case of quarks, through the violation of reflection positivity (signaled by the presence of an inflection
point of the propagator scalar cofactor $\Delta(q^2)$) instead of  achieving an area law for a Wilson loop or a linearly rising potential (criteria which are irrelevant to the question of light-quark
confinement~\cite{Roberts:2012rt}), or satisfying {\it
ad-hoc} criteria involving the ghost sector (which, as already
pointed out above, completely decouples in this regime).

The extension of these quenched lattice results to full QCD, \ie
to a non-Abelian $SU(3)$ theory with the inclusion of dynamical
quarks, has not been extensively pursued, neither in the continuum
nor on the lattice. In the former case, a first
analysis of the effects  on the gluon propagator due to dynamical
quarks  has recently been reported in~\cite{Aguilar:2012rz} within
the so-called PT-BFM (pinch technique-background field method) truncation scheme~\cite{Aguilar:2006gr,Binosi:2007pi,Binosi:2008qk,Binosi:2009qm}. Earlier   related endeavours on the lattice
can be traced back to~\cite{Bowman:2007du} where an ${\cal O}(a^2)$ Symanzik-improved action with $2+1$
staggered fermion flavours was employed, and~\cite{Kamleh:2007ud} where a 
tadpole-improved gauge action with $2$ dynamical overlap fermions was used instead.
However, an independent affirmation of these results by implementing different lattice actions\footnote{Some preliminary
results obtained from simulations with large lattice sizes (far
from the continuum limit) and $\nf=2$ Wilson-Clover fermions have
also been reported in~\cite{Ilgenfritz:2006he}.} as well as their extension for different numbers of flavours, has been a pending issue since then. 

This article provides a comprehensive quantitative study of the
aforementioned Green functions which incorporate the effects
stemming from the presence of dynamical quarks. To this end, we
compute the gluon and ghost two-point Green functions from the
gauge configurations generated by the ETM
collaboration~\cite{Baron:2010bv,Baron:2011sf} for the cases of
 (i) two light degenerate quarks ($\nf=2$) and (ii)
two light and two heavy\footnote{It should be also noticed that
these $2+1+1$ configurations provide a realistic simulation of QCD
below the bottom quark mass threshold, mainly at the momentum
scales which we compute the Green functions for.} ($\nf=2+1+1$)
mass-twisted lattice flavours~\cite{Frezzotti:2000nk}.
Furthermore, we apply our lattice results to carry out an indirect
study of the ghost-gluon form factor (as done for quenched lattice
data in~\cite{Boucaud:2011ug}), by employing a
hybrid approach where the solutions of the ghost SDE are studied
using the gluon propagator determined in our simulations as an input.
Consequently, the natural requirement to reproduce the lattice
ghost dressing function data from the corresponding SDE solution will pin down the ghost-gluon vertex form factor, which will be shown to deviate considerably from its tree-level value.
The constructed SDE solutions then allow us to extrapolate the
lattice ghost data down to the vanishing momentum region and
obtain reliable information on the saturation point of both the
ghost dressing function as well as of the so-called Kugo-Ojima
parameter~\cite{Kugo:1979gm}. Finally, the QCD effective charge,
defined in~\cite{Aguilar:2009nf}, is computed by properly
combining the gluon propagator and the ghost dressing function
with the lattice estimate of the coupling in the so-called Taylor
scheme (\eg see \cite{Boucaud:2008gn}) at a given (large enough)
momentum.

The main results of this
article can be summarized as follows:

\begin{itemize}

\item

The effect of the presence of dynamical quarks on the gluon
propagator $\Delta$ is twofold: a suppression of both the ``swelling''
region at intermediate momenta and the saturation value in the
deep IR (which can be interpreted as the gluon becoming more
massive in the presence of quarks). In addition, one observes that
the more light flavours there are, the bigger the effect is, which
is in accordance with what we would naturally expect. Light
virtual quarks can be copiously produced, thus screening the
interaction and suppressing the very same mechanism which triggers
gluon mass generation. As the fermion mass is increased (at a
fixed flavour number) the effect gets smaller, since the heavier the
fermions, the lesser is the statistical likelihood of their
pair-production. At a sufficiently large value of their mass, they
essentially decouple and the gluon mass generation is practically
insensitive to their presence.  With respect to this
point it should be noticed that our results turn out to be in
agreement with the SDE study reported in~\cite{Aguilar:2012rz}
confirming at the same time the general trend reported in the
earlier lattice studies
of~\cite{Bowman:2007du,Kamleh:2007ud,Ilgenfritz:2006he}.

\item

 On the other hand, the effect on the ghost dressing
function $F$ is much milder and is diametrically opposed to the one
encountered for the gluon case, \ie it consists in a small
increase of the saturation point. This result is also in harmony
with what one would intuitively anticipate. In the SDE for the
ghost, the quark propagator does not enter directly, but only It does so only
through the gluon propagator or via higher loop corrections to the
gluon-ghost vertex. Therefore, it is natural to expect the
influence of dynamical quarks to be less pronounced for the
ghosts.

\item

When the gluon propagator obtained is used as an input in the ghost
SDE,  one finds that the requirement for the SDE solution to match the ghost propagator lattice data
naturally provides a stringent check on the ghost-gluon vertex; specifically, this
exercise will show that this vertex differs significantly from its
tree-level value.

\item

Finally, when all the results are used to form the RG invariant combination $\alpha\Delta F^2$ eventually leading to the QCD effective charge, we observe that, although obviously modifying 
the ultraviolet (UV) parameters controlling the running of the coupling 
and its magnitude, the number of fermions flavours does not affect 
the IR behaviour of this quantity.

\end{itemize}

The paper is  organized as follows:
Section~\ref{general} provides the reader with some of the
technical details of the lattice set-up used for the computation
of the relevant gluon and ghost Green functions. Next, in
Section~\ref{results}, we present the results of the simulations,
emphasizing the differences with respect to the
 quenched results; volume artifacts are also
addressed in some detail. The ghost SDE is then  solved
 in Section~\ref{ghostSDE}, and the effective
coupling  evaluated  in
Section~\ref{sec:eff-coupl}. Finally, we provide the conclusions
  in Section~\ref{concl}.

\section{\label{general}Generalities}

The following section is a reminder of how the ghost and gluon
propagators  are computed from the lattice simulations of gauge
fields for light and heavy mass-twisted lattice flavours. It should
be noticed that these propagators have been obtained (but not presented)
earlier, as a by-product of the computation of the running coupling
in the  momentum subtraction (MOM)  Taylor scheme~\cite{Blossier:2010ky,Blossier:2011tf,Blossier:2012ef}.
These references, which the interested reader is referred to, also contain
relevant details concerning lattice actions, set-ups and the treatment
of artifacts.

In our simulations, the lattice fermion action for the doublet of
light degenerate quarks is  given by \cite{Frezzotti:2003xj}
 \be
S_l =  a^4 \sum_x \overline{\chi}_l(x)
 \left(  D_W[U] + m_{0,l} + i \mu_l \gamma_5 \tau_3 \right)
 \chi_l(x),
\label{eq:tmSl}
 \ee
 whereas, for the heavy doublet, we employ
 \be
S_h  =  a^4 \sum_x \overline{\chi}_h(x) \left( D_W[U] +
 m_{0,h} + i \mu_\sigma \gamma_5 \tau_1 + \mu_\delta \tau_3 \right)
 \chi_h(x),
\label{eq:tmSh}
 \ee
 where $D_W[U]$ stands for the standard massless Wilson Dirac operator.
 In the gauge sector,  the
tree-level Symanzik improved gauge action
(tlSym)~\cite{Weisz:1982zw} is applied for $\nf=2$ and the Iwasaki
improved action~\cite{Iwasaki:1985we,Iwasaki:1996sn} for
$\nf=2+1+1$. In addition to the plaquette term
$U^{1\times1}_{x,\mu,\nu}$,  this formulation of the action  also requires including rectangular
$(1\times2)$ Wilson loops $U^{1\times2}_{x,\mu,\nu}$. For
instance, in the tlSym case, the action reads
\be
    S_g =  \frac{\beta}{3}\sum_x\Bigg\{  b_0\sum_{\substack{
      \mu,\nu=1\\1\leq\mu<\nu}}^4\left[1-\re\,\tr\,(U^{1\times1}_{x,\mu,\nu})\right]
     +
    b_1\sum_{\substack{\mu,\nu=1\\\mu\neq\nu}}^4\left[1
    -\re\,\tr(U^{1\times2}_{x,\mu,\nu})\right]\Bigg\},
  \label{eq:Sg}
\ee
where $\beta \equiv 6 / g_0^2$, $g_0$ is the bare lattice
coupling and one sets $b_1=-1/12$ and $b_0=1-8b_1$ as dictated by
the requirement of continuum limit normalization. Configurations
of the gauge fields generated by the above actions are  next  gauge fixed
to the (minimal) Landau gauge.  This is done through the
minimization of the following functional [of the $SU(3)$ matrices
$U_\mu(x)$]
 \be
F_U[g] = \mbox{\re}\left\{ \sum_x \sum_\mu  \hbox{Tr}\left[1-\frac{1}{N}g(x)U_\mu(x)g^\dagger(x+\mu) \right] \right\},
 \ee
with respect to the gauge group element $g$.

To get as close as possible to the global minimum, we apply a
combination of an over-relaxation algorithm and  Fourier
acceleration, considering the gauge to be fixed when the condition
$|\partial_\mu A_\mu|^2 <10^{-11}$ is fulfilled and the spatial
integral of $A_0$ is constant in time to better than $10^{-6}$.
Evidently, this procedure cannot avoid the possibility that
lattice Gribov copies are  present in the ensemble of gauge fixed
configurations. However, extensive literature in the quenched case
(see for example~\cite{Bogolubsky:2009dc}) shows that such copies
do not seriously affect the qualitative and quantitative behavior
of the Green functions in question.  Given also the
relative large physical volumes simulated, we will proceed under
the working assumption that this feature survives unquenching, as
was also verified in~\cite{Ilgenfritz:2006he}. 

After the lattice configurations have been projected onto the
Landau gauge, one can start calculating the Green functions of
interest. 

To begin with, we consider the gluon propagator.  The
gauge field is defined as
 \be
 A_\mu(x+ \hat \mu/2) = \frac {U_\mu(x) -
 U_\mu^\dagger(x)}{2 i a g_0} - \frac13\, \tr\,\frac{U_\mu(x) -
 U_\mu^\dagger(x)} {2 i a g_0}, \label{amu}
 \ee
 with $\hat \mu$ indicating the unit lattice vector in
the $\mu$ direction. The two-point gluon Green
function is then computed in momentum space through the following
Monte-Carlo average
 \bea
 \Delta^{ab}_{\mu\nu}(q)=\left\langle A_{\mu}^{a}(q)A_{\nu}^{b}(-q)\right\rangle
 = \delta^{ab}\left(\delta_{\mu\nu}-\frac{q_\mu q_\nu}{q^2}\right)\Delta(q^2),
% \left( G^{(2)}\right)^{a_1 a_2}_{\mu_1\mu_2}(p)=
 %\langle A_{\mu_1}^{a_1}(p)A_{\mu_2}^{a_2}(-p) \rangle  \;,
\label{greenG}
 \eea
 with
 \be 
A_\mu^a(q)=\frac 1 2 \,\tr\,\sum_x
 A_\mu(x+ \hat \mu/2)\exp[i q\cdot (x+ \hat \mu/2)]\lambda^a.
 \label{amufour}
 \ee
In the formula above $\lambda^a$ are the Gell-Mann matrices and the trace is evaluated in
color space. 

The Landau gauge ghost propagator can also be computed in terms of
 Monte-Carlo averages of the inverse of the Faddeev-Popov operator,
 \ie
%---
 \bea 
F^{ab}(q^2) = \frac 1 V \ \left\langle \sum_{x,y}
 \exp[iq\cdot(x-y)] \left( M^{-1} \right)^{ab}_{xy} \right\rangle
 =\delta^{ab}\frac{F(q^2)}{q^2},
 \eea
 with $M$
written as a lattice divergence
 \be M(U) =-\frac{1}{N} \nabla
 \cdot \widetilde{D}(U),
 \ee and the operator $\widetilde{D}$
acting on an arbitrary element of the Lie algebra $\eta$ according
to
 \be 
\widetilde{D}(U) \eta(x) = \frac{1}{2}
\left[U_{\mu}(x)\eta(x+\mu) -
 \eta(x)U_\mu(x)+\eta(x+\mu)U_\mu^\dagger-U_\mu^\dagger(x)\eta(x)\right].
 \ee
More details on the lattice procedure for the inversion of the
Faddeev-Popov operator can be found in~\cite{Boucaud:2005gg}.

Next, if we indicate with $\Lambda$ the regularization cutoff (\eg
$\Lambda\equiv a^{-1}(\beta)$ if one specializes to lattice
regularization), one can obtain the renormalized gluon propagator
and ghost dressing function as
 \bea
 \Delta_\s{\mathrm{R}}(q^2,\mu^2)&=&\lim_{\Lambda\to\infty}Z_3^{-1}(\mu^2,\Lambda^2)\Delta(q^2,\Lambda^2),\nonumber \\
 F_\s{\mathrm{R}}(q^2,\mu^2)&=&\lim_{\Lambda\to\infty}\widetilde{Z}_3^{-1}(\mu^2,\Lambda^2)F(q^2,\Lambda^2),
 \eea
where one imposes the standard MOM renormalization  conditions
 \be\label{eq:momcond}
 \Delta_\s{\mathrm{R}}(\mu^2,\mu^2)=1/\mu^2; \qquad
 F_\s{\mathrm{R}}(\mu^2,\mu^2)=1.
 \ee
 When unnecessary, we will refrain from explicitly
indicating the renormalization point dependence of the various
renormalized quantities. 

We conclude this section by commenting briefly on the crucial role
played by the so-called $H(4)$-extrapolation
procedure~\cite{Becirevic:1999uc,Becirevic:1999hj,deSoto:2007ht}, which
have been used  to correct the data for
discretization artifacts (otherwise  plaguing the reliable
determination of $\Delta$ and $F$) due to the breaking of the
$O(4)$ rotational invariance down to the $H(4)$ isometry group.
Specifically, let us observe that the gluon and ghost dressing
functions ($q^2\Delta$ and $F$) are  dimensionless correlation
functions, and therefore general dimensional analysis shows that
they must depend on the (dimensionless) lattice momentum
$a\,{q}_\mu$, where
 \be q_\mu = \frac{2\pi n_\mu}{N_\mu a} \; , \qquad
 n_\mu=0,1,\dots,N_\mu \; ,
 \ee $N_\mu$ being the number of lattice
sites in the $\mu$ direction (in our case, $N_x=N_y=N_z=N_t/2$).
However, if one considers a dimensionless correlator
$Q$ evaluated on the lattice, since $O(4)$ is broken down to
$H(4)$, one has
 \be Q^{\s{\mathrm{latt}}}(a^2q^2,a^2
 \frac{q^{[4]}}{q^2},\cdots) = Q^{\s{\mathrm{latt}}}(a^2 q^2) +
 \left. \frac{\partial Q^{\s{\mathrm{latt}}}}{\partial \left(a^2
 \frac{q^{[4]}}{q^2}\right)} \right|_{a^2 \frac{q^{[4]}}{q^2}=0}
 a^2 \frac{q^{[4]}}{q^2}\,  + \cdots \;, \label{eq:H4}
 \ee where
$q^{[4]}=\sum_\mu q_\mu^4$ is the first $H(4)$-invariant (and the
only one relevant in the ensuing analysis). The
$H(4)$-extrapolation procedure is thought to account properly for
the breaking of $O(4)$ down to $H(4)$ and thus recover the
continuum-limit $O(4)$-invariant result by means of the following
prescription: one first averages over any combination of momenta
being invariant under $H(4)$ (a so-called $H(4)$ orbit); next, one
extrapolates the results towards the continuum limit (where the
effect of $a^2 q^{[4]}$ must vanish) by applying \1eq{eq:H4} to
all the orbits sharing the same value of $q^2$. The only
assumption employed is that the slope coefficient in \1eq{eq:H4}
depends smoothly on $a^2 q^2$.

\section{\label{results}Simulation results}

\begin{table}[!t]
\begin{center}
\begin{tabular}{|c|c|c|c|c|c|c|}
\hline
$\beta$ & $\kappa_{\rm crit}$ & $a \mu_l$ & $a \mu_\sigma$ & $a \mu_\delta$ &
$(L/a)^3\times T/a$ & confs. \\
\hline
\hline
3.90 & 0.161856  & 0.004 &  &  & $24^3\times 64$ & 50 \\
\hline
4.20 &   0.154073 &  0.002 &  &  & $48^3\times 96$ & 50 \\
\hline
\hline
1.95 & 0.161240 & 0.0035 & 0.135 & 0.170  & $48^3\times 96$ & 40 \\
\hline
1.90 &  0.163270 & 0.0040 & 0.150 & 0.190 & $32^3\times 64$ & 50 \\
\hline
\hline
\end{tabular}
\caption{Lattice set-up parameters for the ensembles we used in
this paper: $\kappa_{\rm crit}$ is the critical value for the
standard hopping parameter for the bare untwisted mass; $\mu_l$
stands for the twisted mass for the two degenerated light quarks,
while $\mu_\sigma$ and $\mu_\delta$ define the heavy quarks
twisted masses; the last column indicates the number of gauge
field configurations we used. } \label{tab:set-up}
\end{center}
\end{table}

In this section we describe the outcome of our lattice
simulations. The parameters used are reported in
Table~\ref{tab:set-up}. The physical scale, \ie the lattice size
at any bare coupling $\beta$, has been fixed by European Twisted
Mass Collaboration (ETMC) through chiral fits to lattice
pseudoscalar masses and decay constants. At the physical point,
these are required to take on the values of $f_\pi$ and $m_\pi$
provided by experiments. The bare untwisted mass is tuned to its
critical value by setting the so-called untwisted Partially
Conserved Axial Current (PCAC) mass to zero, so that the
twisted-mass fermions are at {\it maximal twist}. The renormalized
running masses for light and heavy quarks are obtained from the
bare twisted-mass as \bea
\mu_{u,d}(q_0) &=& \frac{a\mu_l} {a(\beta) Z_P(q_0)} \; , \nonumber \\
\mu_{c/s}(q_0) &=& \frac{1}{a(\beta) Z_P(q_0)} \left( a\mu_\sigma
\pm \frac{Z_S(q_0)}{Z_P(q_0)} a\mu_\delta \right),
\eea
where $q_0$ is the renormalization scale. The determination of the
nonperturbative renormalization constants, in particular $Z_P$
and $Z_S$, is the  subject  of an exhaustive computation program
within the framework of ETMC (see for instance~\cite{Baron:2009wt}
for the $\nf=2$ case and \cite{Blossier:2011wk,ETM:2011aa} which
contain some preliminary results for the $\nf=2+1+1$ case). The
degenerate light quark masses we used for the simulations
(Table~\ref{tab:set-up}), range from 20 to 50 [MeV], while the
strange quark is roughly set to 95 [MeV] and the heavy charm to 1.51
[GeV] (in ${\overline{\rm MS}}$ at $q_0=2$ [GeV]). The lightest
pseudoscalar masses for the simulations of Table~\ref{tab:set-up}
range approximately from 270 to 510 [MeV].
The biggest volume simulated corresponds to an asymmetrical
box of roughly $3^3\times6$ [fm$^4$].

\subsection{Gluon sector}

\begin{figure}
%\mbox{}\hspace{-.7cm}
\includegraphics[scale=.9]{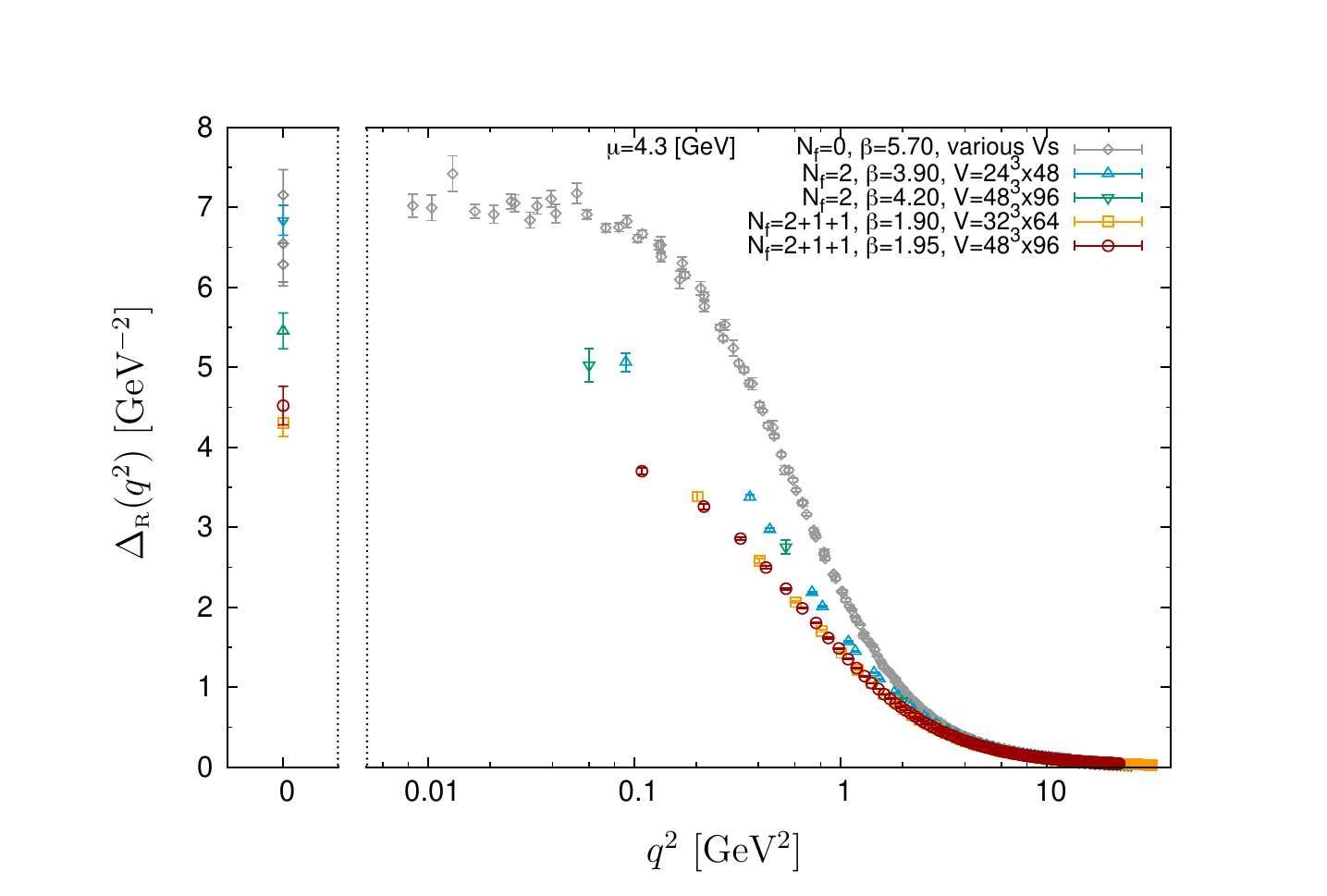}\\
\mbox{}\hspace{-0.73cm}
\includegraphics[scale=.84]{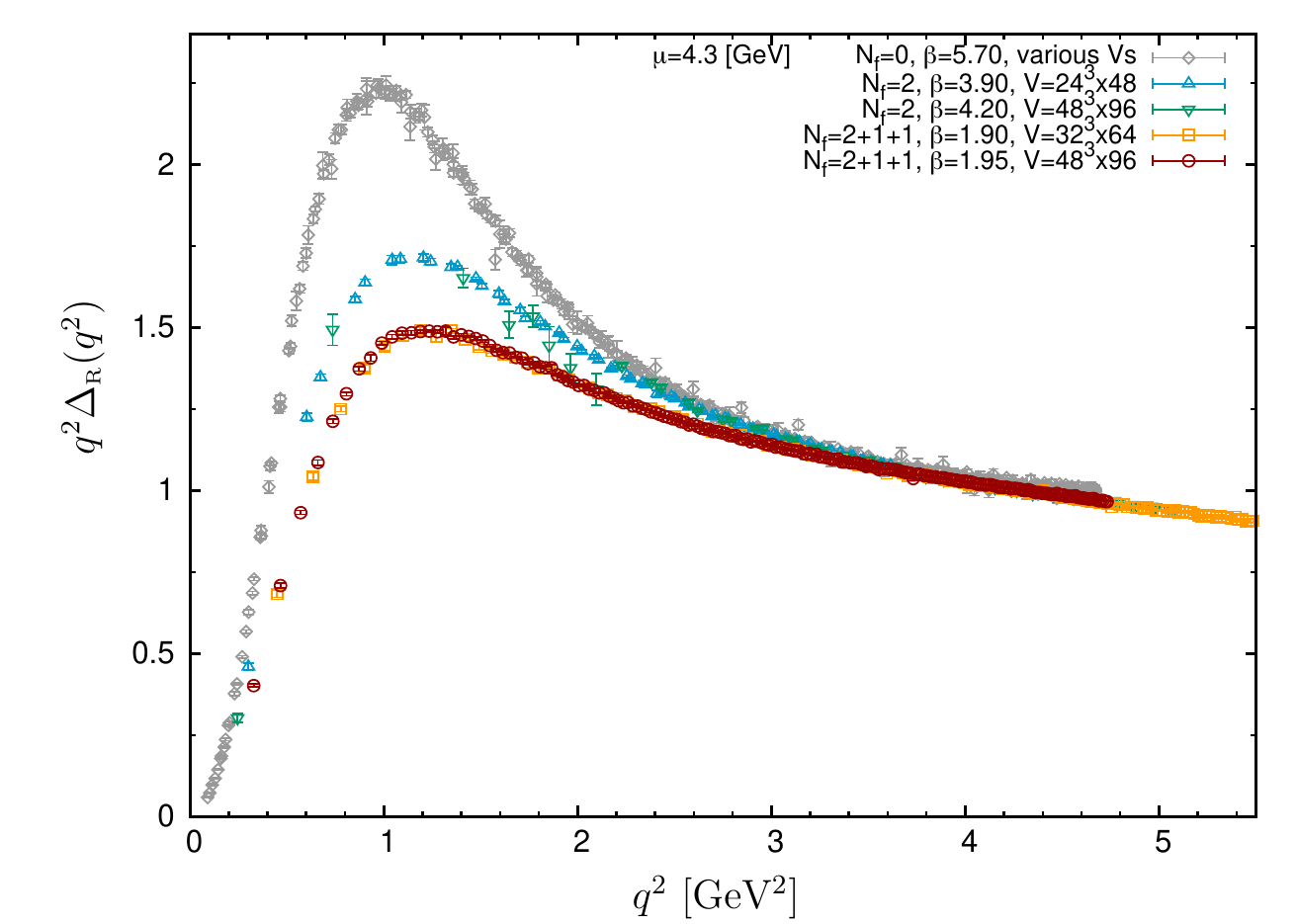}
\caption{\label{gluon}The unquenched gluon propagator (top  panel) and dressing function (bottom
 panel) for $\nf=2$ (two light quarks) and
$\nf=2+1+1$ (two light and two heavy quarks).}
\end{figure}

The results obtained for the gluon propagator and dressing
function for the cases of two light quarks and two light plus two
heavy quarks are plotted\footnote{ If not stated
otherwise we will be setting the renormalization point to be
$\mu=4.3$~GeV}  in \fig{gluon}. As far as the gluon
propagator is concerned (top  panel) one
can clearly see the IR flattening typical of the massive
solutions. However, when compared to the quenched case (shown for
reference by the diamond-shaped gray data points), the propagator shows a less
pronounced ``swelling'' at intermediate momenta and a lower
freezing out value.

\begin{figure}
\includegraphics[scale=.82]{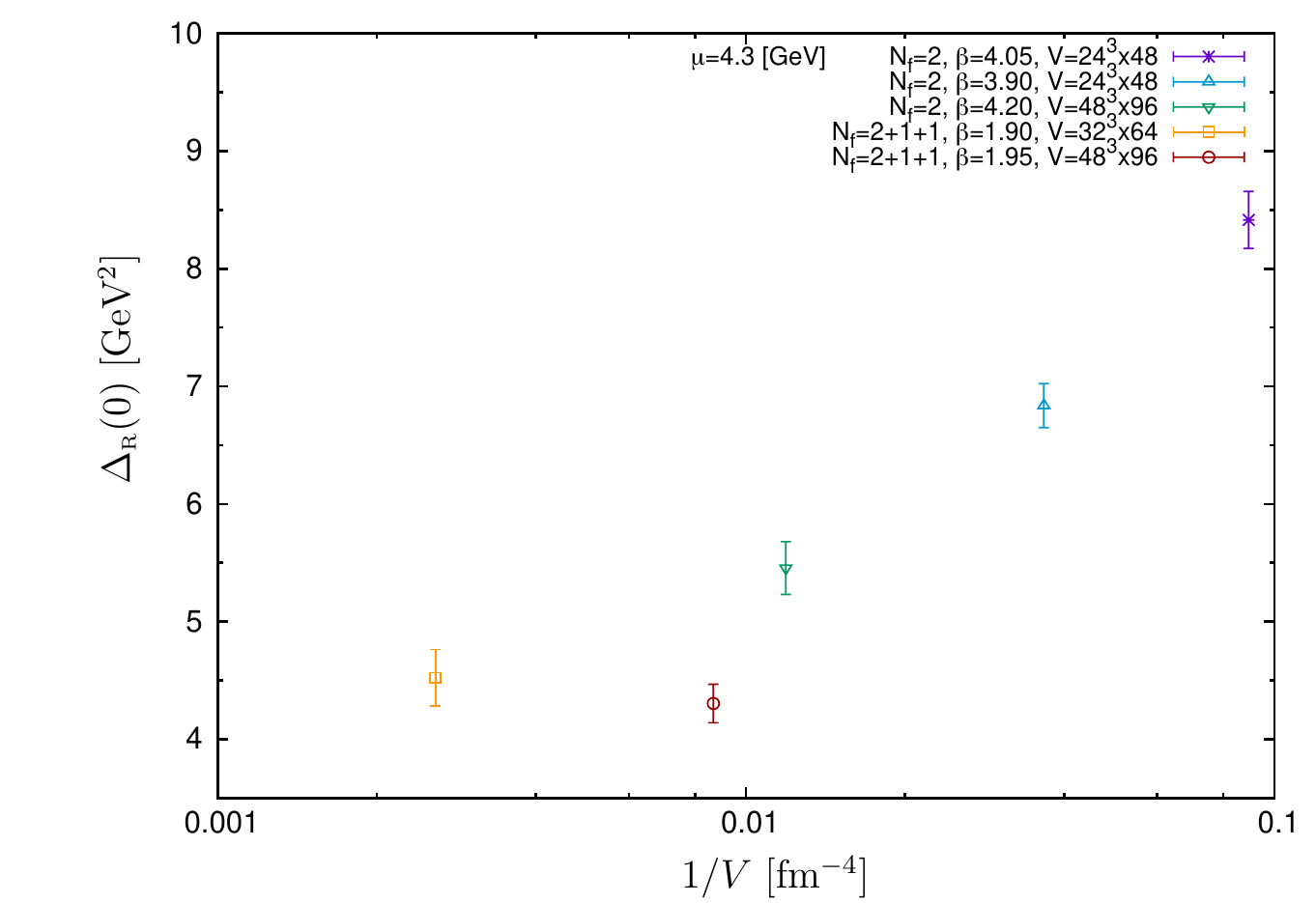}
\caption{\label{volume} The volume dependence of the IR
saturation point of the gluon propagator $\Delta_{\s{\rm R}}(0)$
in our simulations. For the $\nf=2$ case, we include an extra point
corresponding to a simulation on a $24^3\times 48$ lattice, at
$\beta=4.05$ ($\kappa=0.157010$ and $a\mu_l=0.006$).}
\end{figure}

To check the dependence of this latter effect on the lattice
volume, we plot in \fig{volume} the value of $\Delta_{\s{\rm
R}}(0)$ as a function of the inverse of the volume. Though we do
not have enough simulations on large volume lattices to attempt
any continuum extrapolation, it is evident that residual volume
effects are expected to be small when the appropriate simulations
(\ie $\beta=4.20$ for $\nf=2$ and both $\beta=1.95$ and
$\beta=1.90$ for $\nf=2+1+1$) are considered. Furthermore, apart
from the zero-momentum gluon propagator, the results for our two
simulations in both cases appear clearly superimposed in the plots
of Fig.~\ref{gluon}, indicating that volume effects are indeed
under control.

In addition, the quenched simulation can be viewed as an
unquenched counterpart in the limit of infinitely massive
fermions, and the $\nf=2$ results as the limit of the $\nf=2+1+1$
case in the infinite mass limit of the heavy sector. Thus, one can
unambiguously conclude that the presence of dynamical fermions
suppresses the IR saturation point, and renders the gluon heavier.
 Also notice that the suppression tends to subside as
the dynamical fermion mass increases. The decoupling of heavy
fermions has been explicitly shown in the continuum through the
SDE analysis of~\cite{Aguilar:2012rz}, where it was found that the
gluon propagator results for $\nf=2+1$ approach those for $\nf=2$
as the mass of the heavy flavour is increased (see Fig. 17
of~\cite{Aguilar:2012rz}). 

Finally, the concave shape of the
propagator ensures the violation of reflection positivity, thus implying
that the unquenched gluon is also a confined excitation.

The behavior of the dressing function (bottom
 panel) is similar. In this case, the greater the
number of dynamical quarks, the less pronounced the peak at the
intermediate momenta. Analogously, the heavier the quark, the less
the effect it entails on the overall shape of the dressing
function.  These results are in agreement with the SDE
study of~\cite{Aguilar:2012rz}, as well as with the lattice
findings
of~\cite{Bowman:2007du,Kamleh:2007ud,Ilgenfritz:2006he}.

\subsection{Ghost sector}

\begin{figure}
\includegraphics[scale=0.84]{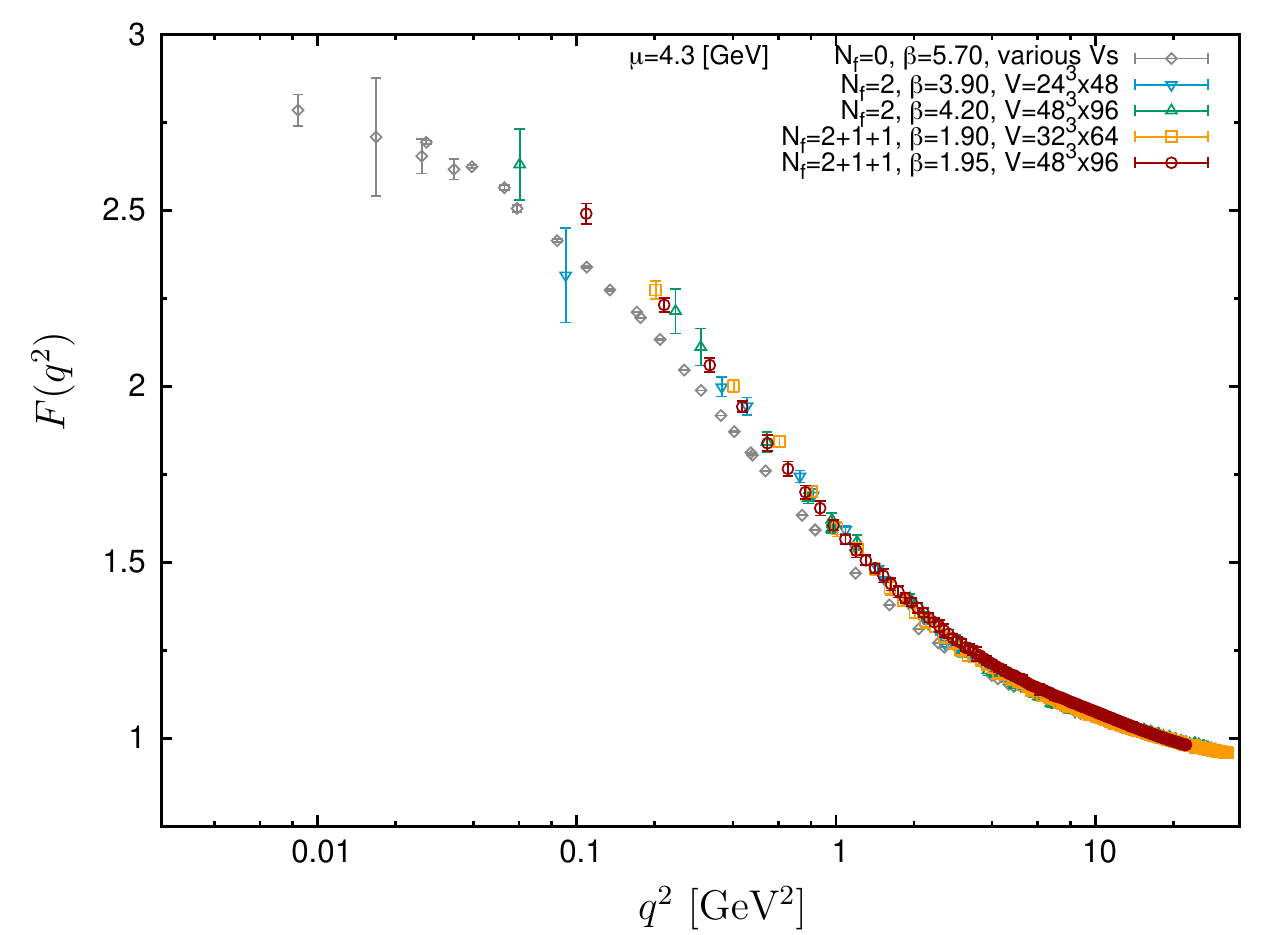}\\
\hspace{-.5cm}\includegraphics[scale=0.84]{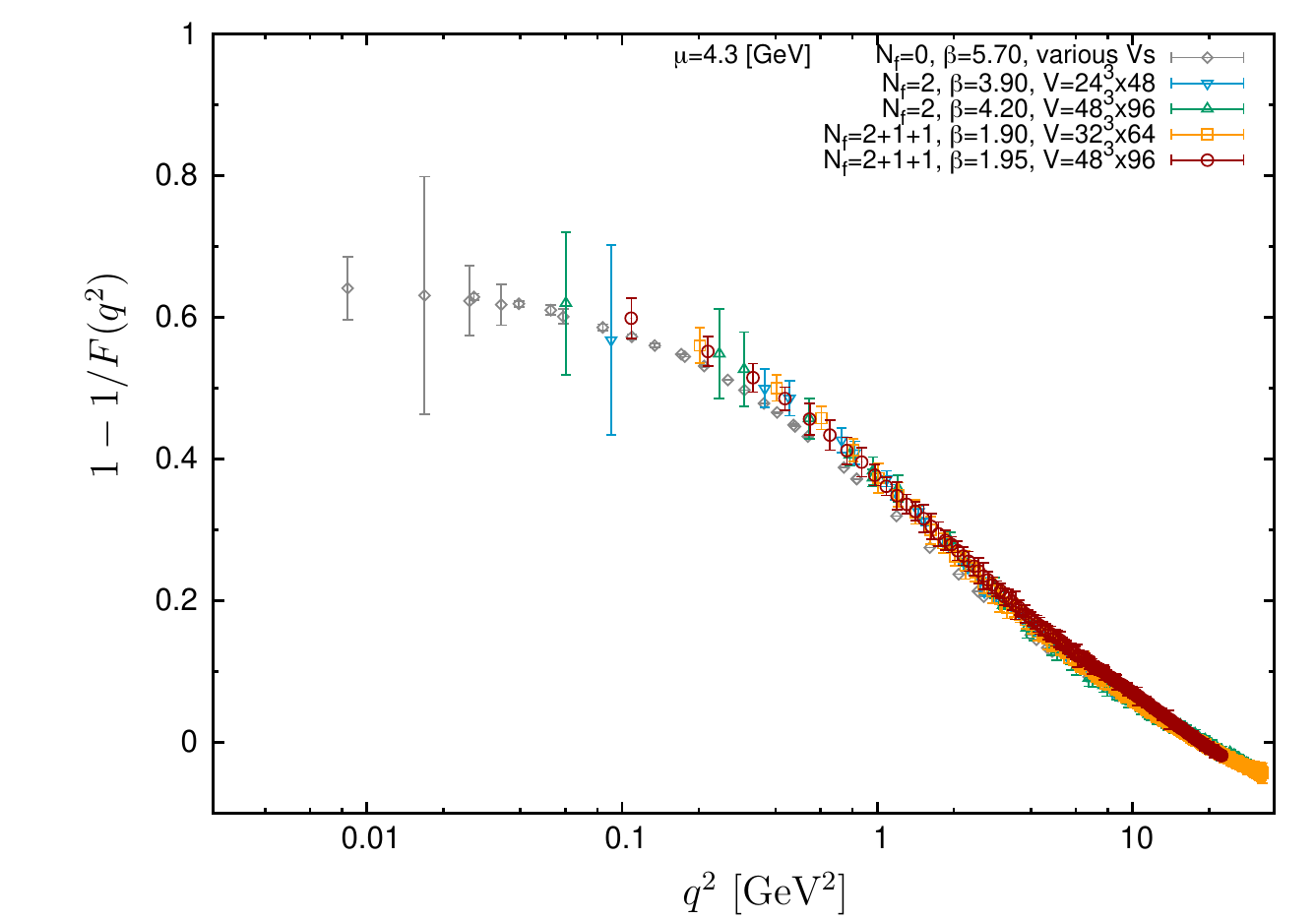}
\caption{\label{ghost}The unquenched ghost dressing function (top  panel) and the (approximate) Kugo-Ojima function (bottom  panel) for $\nf=2$ (two light quarks) and $\nf=2+1+1$ (two light and two heavy quarks).}
\end{figure}

The results for the ghost dressing function are plotted in the
 top  panel of \fig{ghost}. In analogy
with the quenched case, the data do not support a power-like
singular behaviour in the (deep) IR region; rather one finds the
characteristic freezing out feature of the massive
solutions~\cite{Boucaud:2008ji,Boucaud:2008ky}. As one would
expect on the basis of a naive perturbative analysis (there is no
tree level coupling between ghosts and  fermions) the effect of
dynamical quarks on the ghost sector is much milder as compared
to the gluon sector.

The ghost dressing function $F$ can provide valuable information
with respect to the so-called Kugo-Ojima
function~\cite{Kugo:1979gm}. This is due to a powerful identity
dictated by the underlying Becchi-Rouet-Stora-Tyutin (BRST)
symmetry present in the continuum formulation of the theory, which
leads to the relation~\cite{Grassi:2004yq,Aguilar:2009pp} \be
F^{-1}(q^2)=1+G(q^2)+L(q^2), \ee where $G(q^2)$ and $L(q^2)$ are
the form factors of a particular Green function $\Lambda_{\mu \nu}(q)$ that plays a special role in the aformentioned PT-BFM
truncation scheme~\cite{Binosi:2002ez}, with 
 \bea
 \Lambda_{\mu \nu}(q) &=& \delta_{\mu \nu} G(q^2) + \frac{ q_{\mu} \; q_{\nu}}{q^2} L(q^2).
 \eea

The important point here is that $G(q^2)$ coincides (in the Landau gauge) with the
Kugo-Ojima function~\cite{Grassi:2004yq,Aguilar:2009pp}. In
addition, a detailed analysis of the $L(q^2)$ form factor in the
quenched approximation reveals that it is numerically subdominant
in the whole range of momenta when compared to
$G(q^2)$~\cite{Aguilar:2009pp}, and, furthermore, $L(0)=0$.
Since quark effects on $\Lambda_{\mu \nu}(q)$ are suppressed,
either due to their indirect presence in the full gluon and ghost
propagators, or in higher order corrections to the ghost-gluon
kernel (the first one happening at the three-loop level in the
kernel skeleton expansion, and therefore at four loops in
$\Lambda_{\mu \nu}$), one naturally expects the same results to
survive in the unquenched case, thus leaving us with the
approximate relation
 \be G(q^2)\approx F^{-1}(q^2)-1 .
 \ee

In the
 bottom  panel of \fig{ghost} we plot the function $-G(q^2)$ and observe that its
value at origin is practically unchanged when varying the number of
flavours. Clearly the behavior is not dissimilar from the one revealed in quenched simulation,
and the (extrapolated) IR saturation value looks once again far from the critical value 1 predicted by the
scaling type solutions of the SDE and the related Kugo-Ojima confinement criterion. We will return to this issue in the next section.

\section{\label{ghostSDE}Ghost SDE analysis}

In this section we carry out a hybrid analysis combining our
lattice simulation results with  SDE techniques, in a spirit
analogous to what has been reported in~\cite{Boucaud:2008ji}. The
aim is to study the ghost sector in greater detail and, in
particular, gain access to the ghost-gluon vertex form factor(s).
As a welcome byproduct, we will obtain a reliable extrapolation of the
ghost lattice data to the deep IR.

Specifically, let us start by considering the ghost SDE, which can
be recast in the following bare form
\begin{equation}
\label{SD1}
\begin{split}
\frac{1}{F(q^2)} & = 1 + g_0^2 N_c \int \frac{d^4 k}{(2\pi)^4}
 \rule[0cm]{0cm}{0.8cm}
\frac{F(k^2)\Delta((k-q)^2)}{k^2 (k-q)^2}
\left[ \rule[0cm]{0cm}{0.6cm}
\frac{(q\cdot k)^2}{q^2} - k^2
        \right]
\ H_1(k,q),
\end{split}
\end{equation}
where $H_1(k,q)$ is non-longitudinal form factor of the ghost-gluon vertex, parameterized as
\beq
\widetilde{\Gamma}_\nu^{abc}(-k,q;k-q) \ &=& \ i g_0 f^{abc} k_{\nu'}
\widetilde{\Gamma}_{\nu'\nu}(-k,q;k-q) \nonumber \\
&=&
i g_0 f^{abc} \left[k_\nu H_1(k,q) + (k-q)_\nu H_2(k,q) \right],
\label{DefH12}
\eeq
with $k$ and $q$ being the outgoing and incoming ghost
momenta respectively, and $g_0$ the bare coupling constant. As explained in
depth
in~\cite{Boucaud:2008ji,Boucaud:2010gr,RodriguezQuintero:2010wy},
one can first renormalize the ghost and gluon propagators in \eq{SD1},
by using \eq{eq:momcond}, and then apply a subtraction procedure
to deal with the UV singularity of the ghost self-energy integral
to obtain
 \bea
 \frac 1 {F_R(q^2)}  =  1  +
\widetilde{Z}^2_3 Z_3
 \frac{g_0^2}{4\pi} \int\! k^3 {\rm d}k\,
 K(k,q) H_1^{\rm bare}(k,q) F_R(k^2) ,
\label{eq:numDSE}
 \eea
where
 \bea
  K(k,q)  =  - \frac 1 {\pi^2} \ \int_0^\pi\!
 \sin^4{\theta}\, {\rm d}\theta  \left[ \frac{\Delta_R((k-q)^2)}{(k-q)^2} -
 \frac{\Delta_R((k-p)^2)}{(k-p)^2} \right].
\label{eq:deckernel}
 \eea
The renormalization point, $\mu^2$, is implicitly present as an
argument for all the renormalized quantities. In obtaining
\eq{eq:numDSE}, the subtraction procedure is applied for \eq{SD1}
evaluated at the two momenta $k$ and $p$, both being parallel and
such that $p^2=\mu^2$. $H_1$ in \eq{eq:numDSE} is a bare but
finite~\cite{Taylor:1971ff} quantity which needs no
renormalization while, in front of the integral, the
renormalization constants and the bare coupling especially appear
in the right combination to cancel the cut-off dependence and give
the MOM Taylor scheme coupling (see \eg \cite{Boucaud:2008gn}),
\be \alpha_\s{\rm T}(\mu^2)  =  \frac{g^2_0}{4\pi}
\widetilde{Z}_3^2(\mu^2) Z_3(\mu^2). \label{alphaT} \ee This
coupling  $\alpha_\s{\rm T}(\mu^2)$ for $\nf=0,\ 2$ and $2+1+1$
can be determined from lattice data (see,
\eg~\cite{Boucaud:2008gn,Blossier:2010ky,Blossier:2011tf,Blossier:2012ef}).
In order to solve the ghost SDE in isolation (\ie without coupling
it to the much more complicated gluon SDE), one can use the just
determined lattice gluon propagator $\Delta_\s{\rm R}$ as an input
for the equation, thus fully determining the
kernel~\noeq{eq:deckernel}. Now the only unknown term present in
the equation is the ghost form factor $H_1$; clearly the solutions
to the ghost SDE will describe the lattice data with a better or
worse agreement depending on our ability to model this form
factor~\cite{Boucaud:2011ug,Boucaud:2011eh}.%,DORQ:2012}.

Through the analysis of the quenched lattice data, it was shown
in~\cite{Boucaud:2008ji} that the solutions of~\1eq{eq:numDSE}
grossly underestimate (by a factor of at least 2) the lattice data
if one uses the tree-level value $H_1=1$ for the ghost-gluon form
factor. A constant does indeed do a better
job~\cite{Boucaud:2008ji} but does not allow for a precise
description of the deep IR behavior of the
function~\cite{Boucaud:2011ug}. %,DORQ:2012}. 
This implies that a good
description of the (quenched) ghost dressing lattice data calls
for a ghost-gluon form factor with a non-trivial kinematical
structure. Using the knowledge derived from the OPE analysis
of~\cite{Boucaud:2011ug}, coupled with the current lattice data on 
the (Landau gauge) ghost-gluon vertex~\cite{Maas:2006qw}, one can
parameterize this form factor as\footnote{\1eq{eq:numDSE} requires
to be solved with the full vertex $H_1(k,q)$, which is modelled
in~\cite{Boucaud:2011ug}; however it can be shown that $H_1(k,q)
\simeq H_1(k,0)$ is a good approximation to obtain the ghost
dressing in the IR momentum region~\cite{Dudal:2012zx}.} \be H_1(k,0)
=  H_1^0 \left[ 1 + \frac{N_\s{C} g^2\langle A^2 \rangle}{4
(N_\s{C}^2-1)} \frac{k^2}{k^4+m_{\rm IR}^4} \right]+ \left( 1 -
H_1^0 \right) \frac{w^4}{w^4 + k^4}. \label{eq:fullvertex} \ee
\begin{figure}
\includegraphics[scale=0.84]{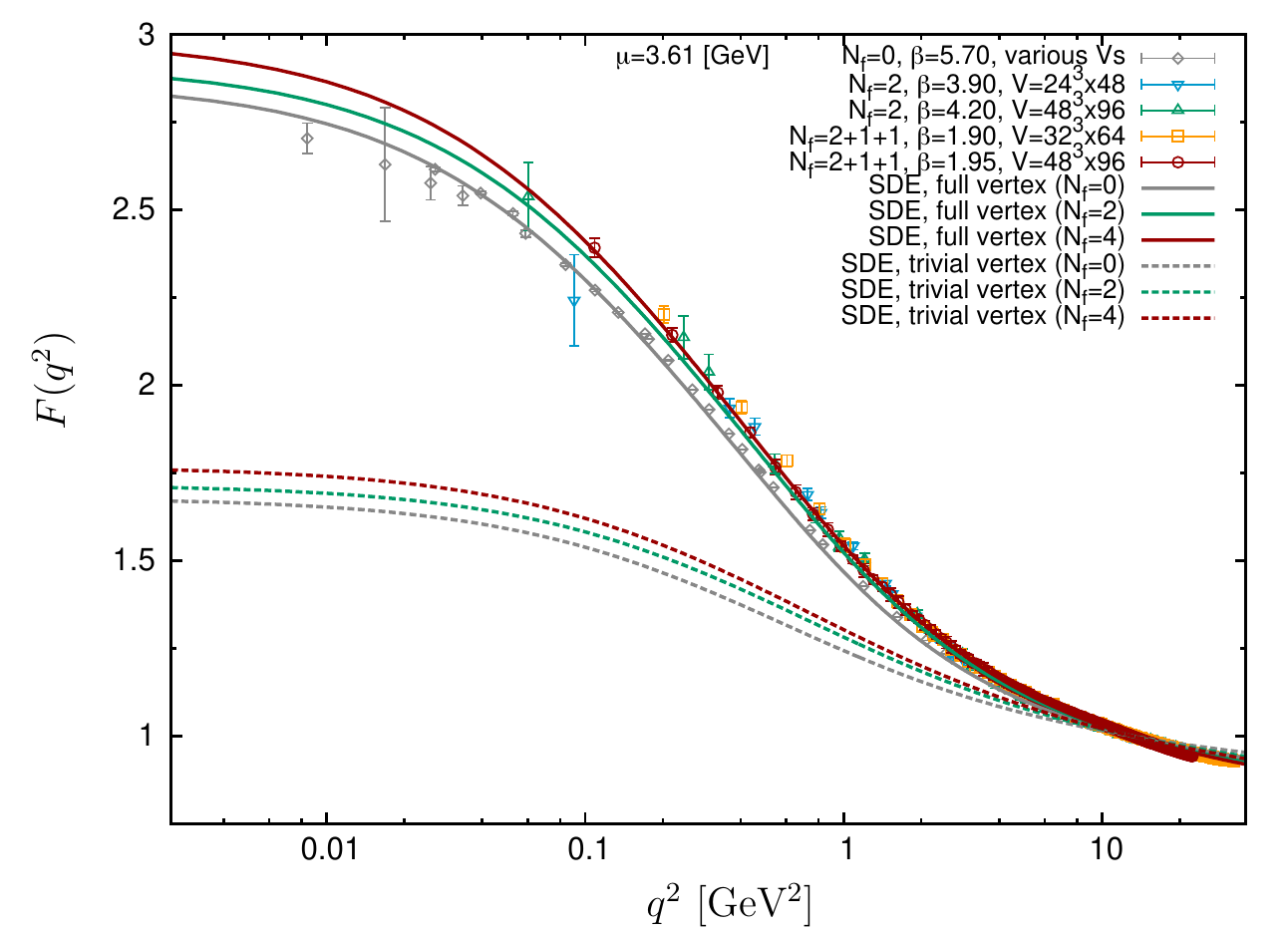}\\
\hspace{-0.2cm}\includegraphics[scale=0.84]{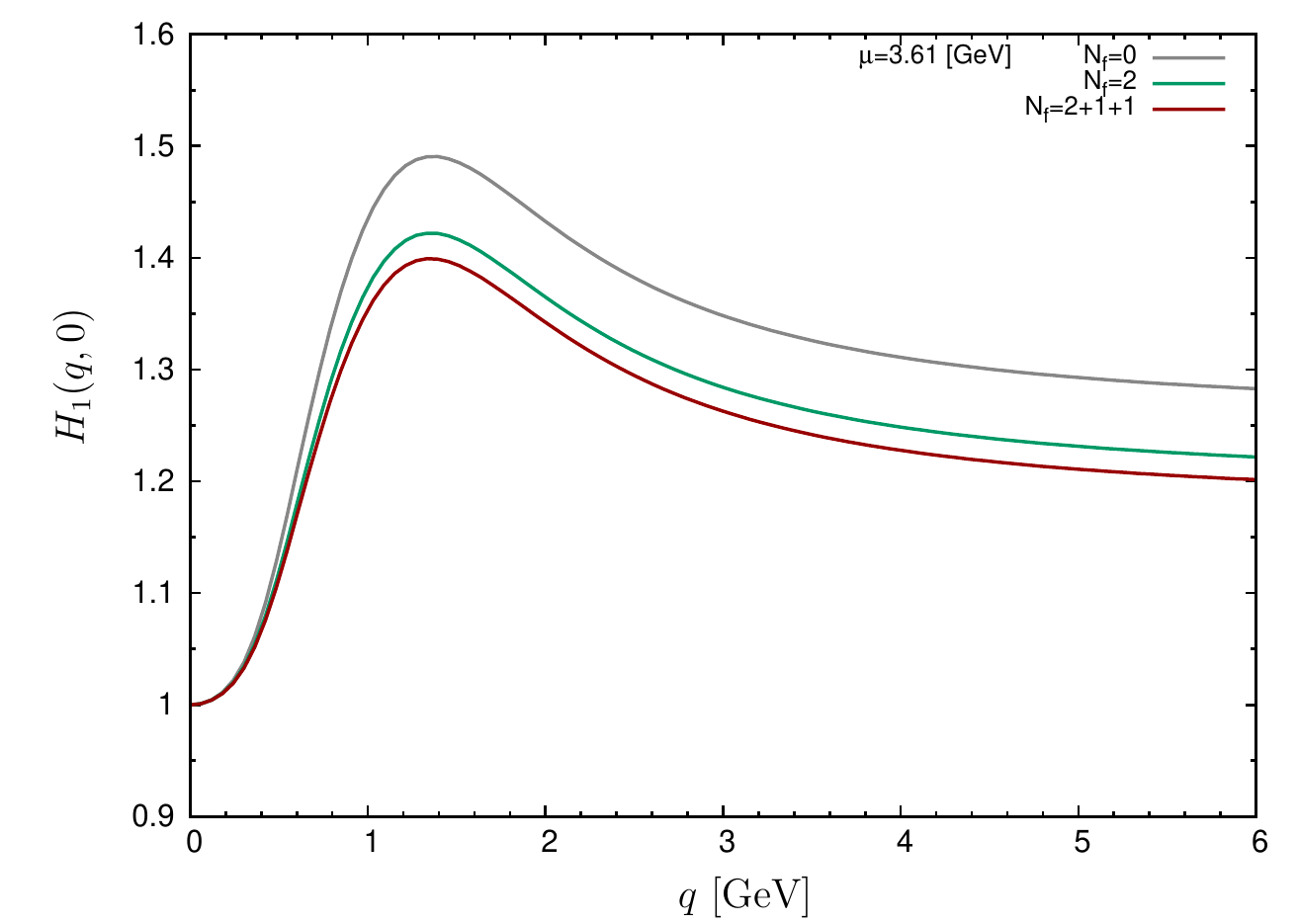}
\caption{\label{ghSDE} (Top  panel) The
ghost dressing function obtained from the solution of the ghost
SDE~\noeq{eq:numDSE} with the lattice gluon propagator as an input
and $\alpha_\s{\rm T}=0.25,\ 0.32,\ 0.37$, respectively, for
$\nf=0,\ 2$, and $2+1+1$ at $\mu=3.61$ [GeV]. Dashed lines
correspond to solutions for the tree-level $H_1(q,0)=1$, while
continuous lines to the inclusion of the full form
factor~\noeq{eq:fullvertex}. The latter is also shown in the
 bottom  panel. The values of the
parameters used to integrate the ghost SDE are~:  $g^2\langle A^2
\rangle=7$ GeV$^2$, $m_{\rm IR}=1.3$ GeV and $w=0.65$ GeV (the
same IR ones for the three cases). Moreover, $H_1^0=1.26$, $1.20$,
$1.18$ for $\nf=0$, $2$ and $2+1+1$, respectively.}
\end{figure}
\begin{figure}
\includegraphics[scale=0.84]{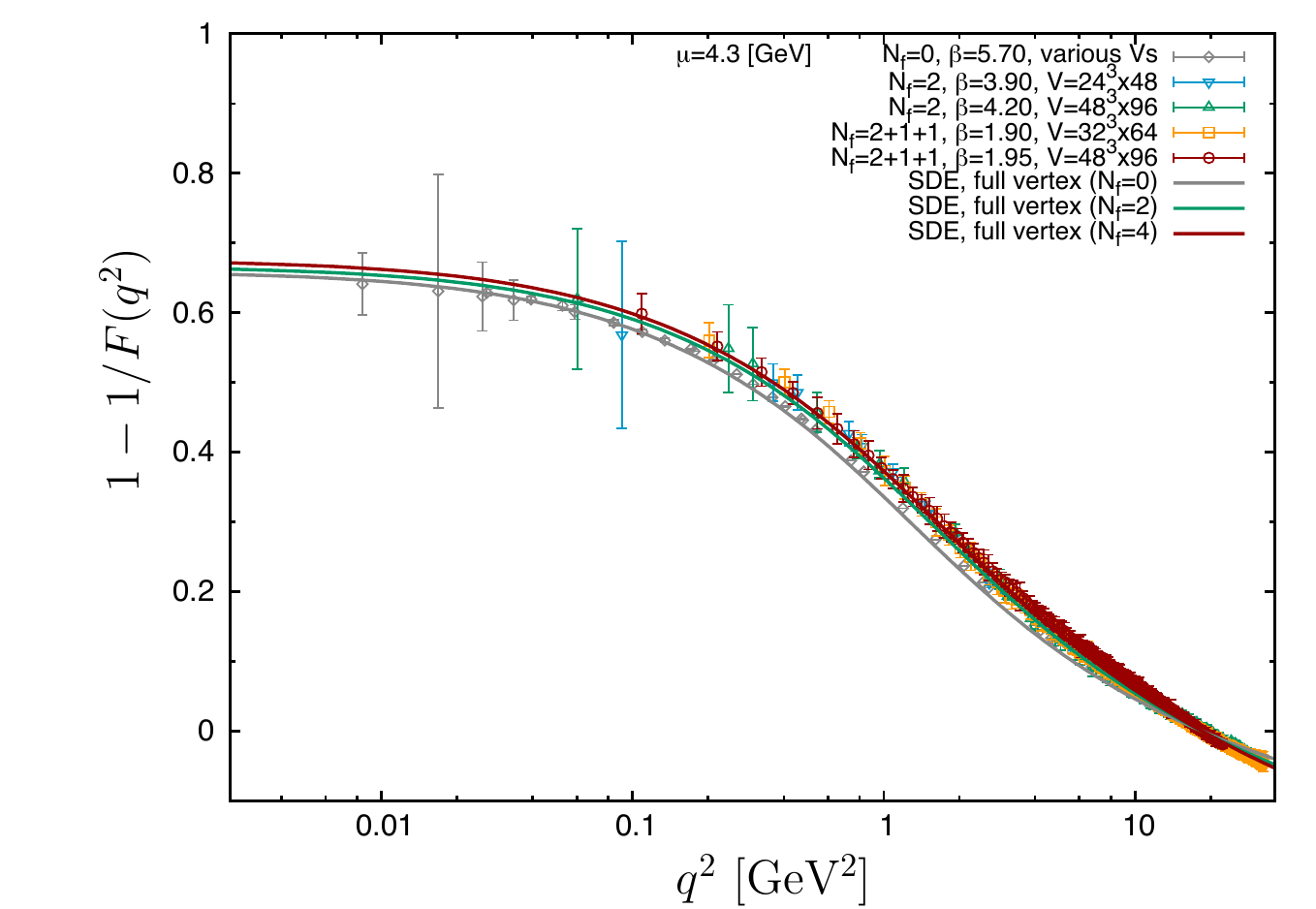}
\caption{\label{ghSDE-G} Extrapolation of the (approximate) Kugo-Ojima function
in the deep IR. The ghost dressing function $F$ employed in this
plot is generated by solving the ghost SDE.}
\end{figure}
Estimates for the gluon condensate\footnote{The very notion of condensate have  been recently questioned in~\cite{Brodsky:2010xf,Chang:2011mu}.  In particular, according to the new perspective suggested there, our gluon condensate $g^2 \langle A^2 \rangle$ should be understood as a mass-scale parameter related to the local 
operator $A^2$ in the OPE expansion of the gluon Green functions.} $g^2 \langle A^2 \rangle$, and
the IR mass scale $m_{\rm IR}$, can be obtained from lattice data
and OPE analysis~\cite{Boucaud:2011eh}; the constants $H_1^0$ and
$w$ (introduced in order to guarantee that $H_1(0,0)=1$, as
suggested by current lattice data~\cite{Maas:2006qw}) can be
adjusted so that the solutions~\1eq{eq:numDSE} match the
corresponding lattice data as closely as possible. 
%The latter is being deeply investigated for quenched lattice data in ref.~\cite{DORQ:2012}.

The solutions of the ghost SDE~\noeq{eq:numDSE} following the
procedure just illustrated are presented in~\fig{ghSDE}.
 In the top panel of the figure one can clearly see
that, similarly to the quenched case,  a tree-level
value for $H_1$ does not give solutions which can describe the
lattice data. However, once the kinematically non-trivial
expression,~\1eq{eq:fullvertex} (bottom 
panel of the same figure), is included in the equation, one obtains
an excellent agreement with the data. Therefore, effectively, the
curves for $H_1(q,0)$ represent a genuine prediction of our
analysis. It would be interesting to confirm or refute this
prediction through direct lattice calculations of the ghost-gluon
three-point function.

The good agreement between the SDE solutions and the lattice
data allows for an extrapolation of the latter towards the deep IR
region, where one observes a very small increment of the
saturation point (monotonic with $\nf$). This is particularly
useful when scrutinizing the Kugo-Ojima function as shown in
\fig{ghSDE-G}, which clearly depicts that the saturation point of
the function is practically insensitive to the inclusion of
dynamical fermions.

\section{\label{sec:eff-coupl}Effective coupling}

The results obtained for the gluon  and ghost two-point functions
allow us to extract the running of the full QCD effective charge
for a wide range of physical momenta, and in particular in the
deep IR region which is  evidently 
inaccessible to perturbation theory.

To begin with, let us recall that the QCD effective charge is
defined, among practitioners, in primarily two different ways: the
first one (to be denoted by $\alpha_\s{\rm PT}$) is obtained
within the framework of the pinch
technique~\cite{Cornwall:1981zr,Binosi:2009qm} and represents the
most direct generalization of the familiar QED effective charge
concept to a non-Abelian setting; the second one (to be denoted by
${\alpha}_\s{\rm gh}$) corresponds to the
non-perturbative generalization of the strong coupling in the
Taylor scheme mentioned
before~\cite{Blossier:2010ky,Blossier:2011tf,Blossier:2012ef}.

\begin{figure}
\includegraphics[scale=0.9]{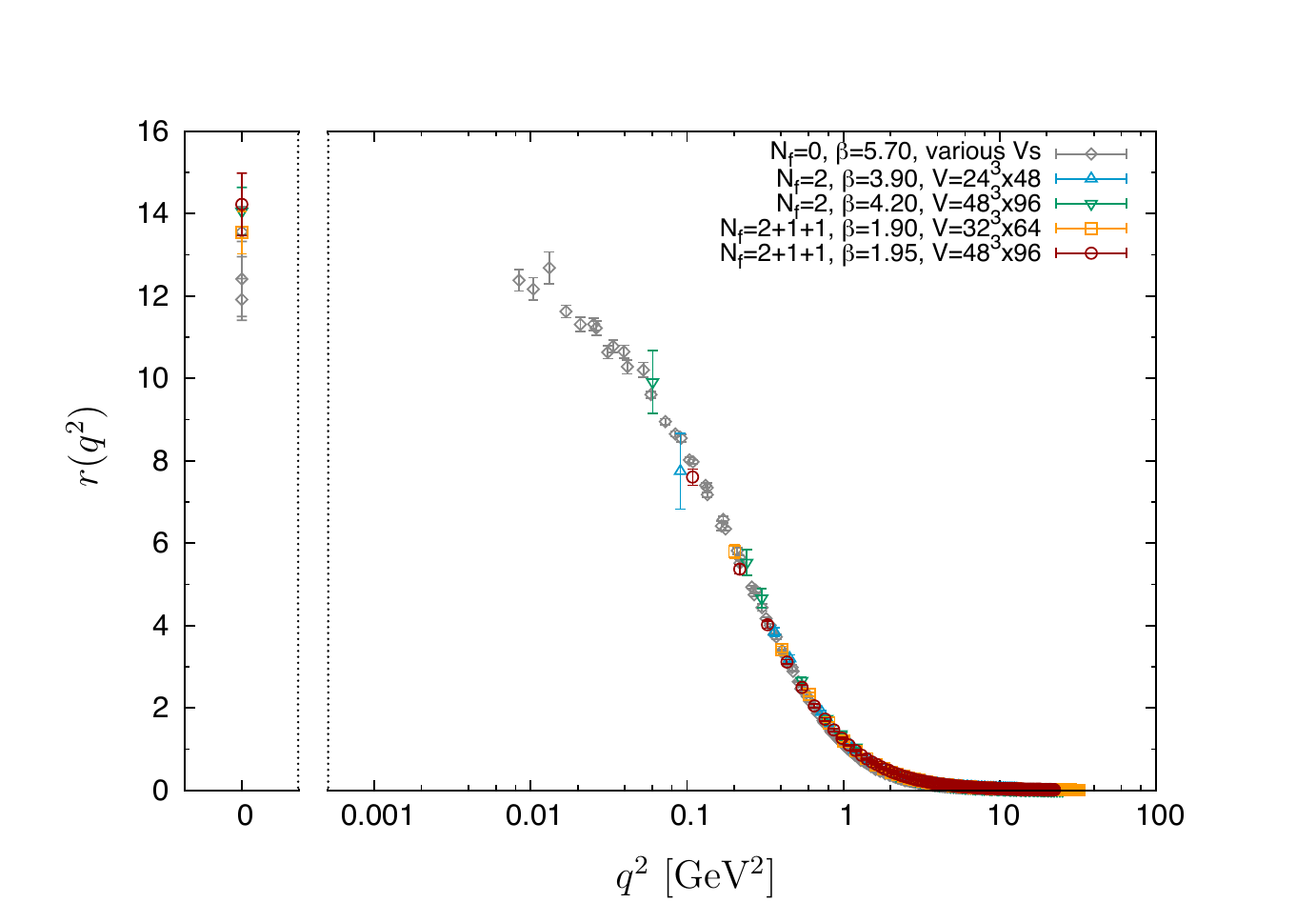}
\caption{\label{RGI} The RG invariant quantity $r(q^2)$ defined in~\1eq{RGIcomb} for the different values of flavours $\nf$. Errors for the $\nf=0$ case are underestimated, since we have used the ghost dressing function obtained from the SDE for constructing the effective charge.  Notice the absence of any
flavour dependence below the 1 GeV region.}
\end{figure}

The construction of either effective charges proceeds through the identification of a suitable RG invariant combination. Before identifying this quantity however, let us observe that
though the effective couplings $\alpha_\s{\rm PT}$ and  ${\alpha}_{\rm gh}$ have a rather distinct theoretical origin
and status, it turns out that, in the Landau gauge, they are
related through the equation~\cite{Aguilar:2009nf}
 \be
{\alpha}_{\rm
gh}(q^2)=\left[1+\frac{L(q^2)}{1+G(q^2)}\right]^{-2}\alpha_\s{\rm
PT}(q^2);
 \ee
evidently, in the approximation $L(q^2)\approx0$, used throughout this paper, the two definitions coincide, and one has
${\alpha}_\s{\rm PT}(q^2)\equiv\alpha_{\rm gh}(q^2)\equiv
\overline{\alpha}(q^2)$. This implies also that one can choose as the RG invariant combination 
\be
r(q^2)=\alpha_\s{\rm
T}(\mu^2)\Delta_\s{\rm R}(q^2,\mu^2)F^2_\s{\rm R}(q^2,\mu^2),
\label{RGIcomb} 
\ee 
which can be readily obtained from the data presented so far [$\alpha_\s{\rm T}$ is given in \eq{alphaT}]. The quantity $r(q^2)$ defined above is constructed in \fig{RGI} for 
different number of flavours $N_f$; notice that for calculating the freezing out point  $r(0)$ the value of $F_\s{\rm R}(0)$ has been extrapolated from the SDE results for the ghost dressing obtained in the previous section.

A most salient feature of this plot is the absence (within the errors) of any flavour dependence in the IR region, more
precisely starting from $q^2\lesssim 1$ GeV$^2$. Indeed, one observes that the flavour effects which control the behavior of the UV parameters 
of the theory (\eg the $\beta$-function coefficients, $\Lambda_\s{\rm QCD}$ and $\langle A^2\rangle$), combine in such a way 
that, when the RG invariant combination $r(q^2)$ is formed, no net flavour dependence survives in the IR .

Since, modulo an overall dimensionful factor to render it dimensionless, $r(q^2)$ coincides with the effective coupling, the origin of this independence 
can be understood by recalling the reason for the $\nf$ dependence of the running coupling in the 
UV (it should also be noticed that the invariant combination $r(q^2)$ is related with the UV coupling defined through the ghost-gluon vertex 
in Taylor scheme by nothing but a factor $q^2$). In this case the bigger the physical momenta $q^2$, the more channels open up for 
the production of quark anti-quark virtual pairs (so that every time a channels opens, the
coupling receives a ``kick'' and goes up). However, as soon as
$q^2$ drops below a certain threshold, no energy will be available
to produce any virtual pairs (not even gluons when $q^2_0<4m_0^2$)
so that the residual running of the coupling below this value is
completely dominated by the  IR mass scale introduced
when defining the effective charge.

Coming to this specific point, it turns out that~\cite{Aguilar:2009nf} one can construct from $r(q^2)$ the dimensionless effective coupling $\overline{\alpha}(q^2)$ by pulling out the
inverse propagator factor \mbox{$q^2+m^2(q^2)$}, \ie
\be
\overline{\alpha}(q^2)=\left[q^2+m^2(q^2)\right]r(q^2),
\label{effch}
\ee
which leads to an IR saturating coupling. Notice that  this definition is valid
for both massive and the (already ruled out) scaling
solutions (in which case one would have to set $m^2(q^2)=0$);
since in the latter case $\Delta(0)\rightarrow 0$ and
$F(0)\rightarrow\infty$, the effective coupling does not distinguish between the two solutions\footnote{As
explained in detail in~\cite{Aguilar:2009nf}, in the presence of
an IR saturating propagator, one should not insist in pulling out
in front of the effective coupling a simple $q^2$ factor.
Otherwise, one would end up with a completely unphysical coupling,
namely the one that vanishes in the IR, where QCD is supposed to
be a strongly coupled theory.}. 

\begin{figure}
\includegraphics[scale=0.9]{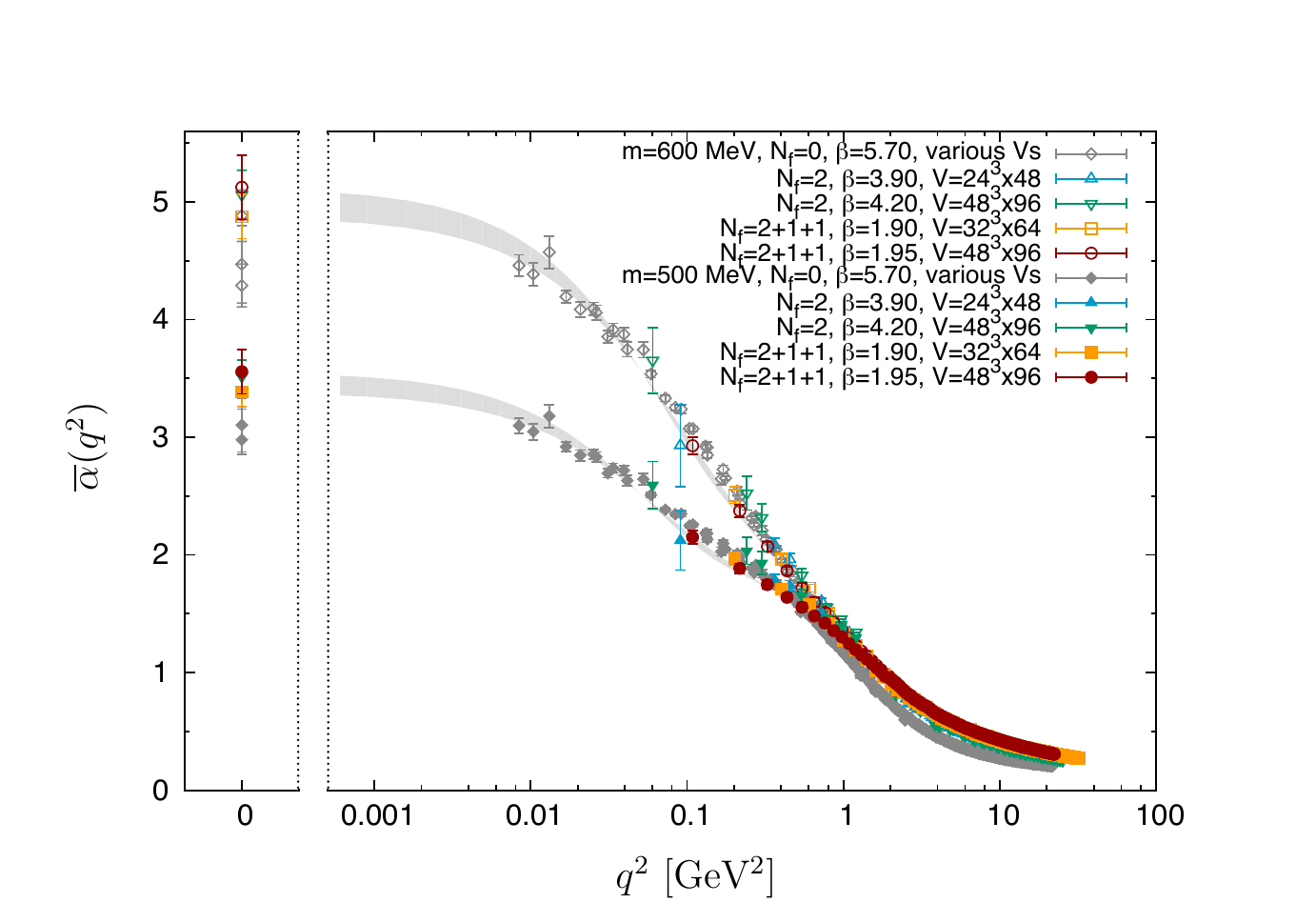}
\caption{\label{eff-coupl} The effective charge
$\overline{\alpha}(q^2)$ defined in~\1eq{effch} for two different
values of the IR gluon mass: $m_0=500$ MeV (solid symbols) and $m_0=600$ MeV (open symbols). The gray band in the background (meant to guide the eye)  has been obtained from a continuum extrapolation of the $\nf=2+1+1$ data.}
\end{figure}

As a last step, we need to specify the $q^2$ running of the  dynamical mass  $m^2(q^2)$. We will consider here  the simplified setting of~\cite{Aguilar:2009nf,Aguilar:2010gm} under which the mass obeys a power law running
\be
m^2(q^2)=\frac{m_0^4}{q^2+m_0^2}; \qquad m_0 \equiv m(0),
\label{run}
\ee
and for $m_0$ one considers the  representative values $m_0=500 - 600$ MeV, consistent with a variety of phenomenological studies. The resulting effective charge is plotted in \fig{eff-coupl}.

We hasten to emphasize that this is only a toy model, and one should take into account that in reality $m_0$ differs for different $\nf$, as clearly 
seen in the top panel of~\fig{gluon}. However, inserting directly in~\1eq{run} the 
saturation value $m_0=\Delta_\s{\rm R}^{-1}(0,\mu^2)$ obtained from our simulations for different number of flavours $\nf$, 
breaks the RG invariance in zero of~\1eq{effch}\footnote{One could in principle choose different phenomenological values of $m_0$ 
for different values of $\nf$ but the result would not be that different from what is seen in \fig{eff-coupl}.}.
We are evidently in need of better tools for extracting reliable (RG invariant) information about the saturation point of the coupling 
(and probably more data as well in the low momentum region); this issue clearly deserves a separate study. 

\section{\label{concl}Conclusions}

In this paper, we have carried out a systematic and comprehensive
analysis of the gluon and ghost two-point functions in (Landau
gauge) full lattice QCD. 

The configurations used include two light
and two light plus two heavy twisted mass fermions with masses
between 20$-$50 [MeV] for the light quarks, 95 [MeV] for the
strange quark and 1.51 [GeV] for the charm quark (in
$\overline{\rm MS}$ scheme at a renormalization scale of 2 [GeV]).
The mass of the lightest pseudoscalar turns out to be between the
range of 270 and 510 [MeV]. As this value does not lie too far
from the physical pion mass, it increases our confidence in the flavor
physics effects reported in this article. Moreover, simulations on
lattices with up to $48^3\times96$ points, with $\beta=3.90$ and
$4.20$ for $\nf=2$ and $\beta=1.90$ and $1.95$ for $\nf=2+1+1$,
allow us to reach momenta down to $q\simeq 300$ [MeV], keeping the
volume effects under control.

   Our analysis demonstrates that in the intermediate and low
momentum region, the gluon propagator lessens with the increase in
the number of dynamical quarks, whereas, the ghost dressing
function is enhanced, albeit only slightly. In addition, the
heavier a species of fermions, the smaller in extent is its effect
on the suppression of the gluon propagator. With a heavier enough
mass, which prevents its virtual pair production, the fermion
fails to screen the interaction and gets decoupled from the gluon
dynamics altogether.

%. In a particular, as a naive perturbative analysis would suggest,
%these effects can be clearly identified in the gluon sector (where
%our results are also in perfect agreement with analogous SDE
%study), whereas the ghost sector is left practically unchanged.

When all the pieces of data are put together to construct the
effective QCD running coupling, $\overline{\alpha}(q^2)$, we observe the
behavior anticipated from the massive decoupling solutions,
namely, a monotonic approach to an IR fixed point. Furthermore, we
find that below $q\simeq1$ [GeV], this quantity is not directly
affected by the variation in the number of dynamical fermion
flavours. However, considering that an IR gluon mass is introduced
while defining the effective running coupling, there is indirect
dependence on $\nf$ via this mass scale.

Making the most of the lattice results for the gluon and ghost
propagators, we present a self-consistent analysis of the ghost
2-point function and extract the unquenched ghost-gluon form factor $H_1$. 
This is a genuine prediction of the SDE study presented in Sect.~\ref{ghostSDE}, 
which should be confirmed (or refuted) through direct lattice studies of 
the ghost-gluon vertex.

Though the data presented here have been obtained for an arbitrary
gauge copy selected through a gauge fixing algorithm using a
combination of over-relaxation and Fourier acceleration, we do not
expect that the presence of Gribov copies to alter the conclusions
in any significant way, given also that their effect is expected
to weaken significantly for physical volumes as large as the ones
considered here.

\acknowledgments

A.B. wishes to acknowledge the financial grants CONACyT Project
46614-F and Coordinaci\'on de la Investigaci\'on Cient\'{\i}fica
(CIC) Project No. 4.10 and U.\,S.\ Department of Energy, Office of Nuclear Physics, contract no.~DE-AC02-06CH11357. 
The work of M.C. is supported by the AuroraScience project, which is funded jointly by the Provincia Autonoma di Trento (PAT) and the Istituto Nazionale di Fisica Nucleare (INFN). Part of the calculations were performed using the Aurora Supercomputer at the Fondazione Bruno Kessler (Trento). J.~R.-Q.~is indebeted to O.~P\`ene, Ph. Boucaud, B. Blossier, K. Petrov and C. D. Roberts for fruitful discussions and very helpful comments, and also acknowledges the Spanish MICINN for the support by the research project FPA2011-23781 and ``Junta de Andalucia'' by P07FQM02962.

%\bibliographystyle{ieeetr}
%\bibliography{total}

\end{document}